\documentclass{article}

 \usepackage[preprint]{neurips_2025}

\usepackage[utf8]{inputenc} % allow utf-8 input
\usepackage[T1]{fontenc}    % use 8-bit T1 fonts
\usepackage{hyperref}       % hyperlinks
\usepackage{url}            % simple URL typesetting
\usepackage{booktabs}       % professional-quality tables
\usepackage{amsfonts}       % blackboard math symbols
\usepackage{nicefrac}       % compact symbols for 1/2, etc.
\usepackage{microtype}      % microtypography
\usepackage{xcolor}         % colors
\usepackage{graphicx}
\usepackage{float}
\usepackage{amsmath,amssymb}

\title{Earthquake Forecasting with ETAS.inlabru}

\author{%
  Ziwen Zhong
}

\begin{document}

\maketitle

\begin{abstract}
The ETAS models are currently the most popular in the field of earthquake forecasting. The MCMC method is time-consuming and limited by parameter correlation while bringing parameter uncertainty. The INLA-based method “inlabru” solves these problems and performs better at Bayesian inference.

The report introduces the composition of the ETAS model, then provides the model's log-likelihood and approximates it using Taylor expansion and binning strategies. We also present the general procedure of Bayesian inference in inlabru.

The report follows three experiments. The first one explores the effect of fixing one parameter at its actual or wrong values on the posterior distribution of other parameters. We found that $\alpha$ and $K$ have an apparent mutual influence relationship. At the same time, fixing $\alpha$ or $K$ to its actual value can reduce the model fitting time by more than half.

The second experiment compares normalised inter-event-time distribution on real data and synthetic catalogues. The distributions of normalised inter-event-time of real data and synthetic catalogues are consistent. Compared with Exp(1), they have more short and long inter-event-time, indicating the existence of clustering. Change on $\mu$ and $p$ will influence the inter-event-time distribution.

In the last one, we use events before the mainshock to predict events ten weeks after the mainshock. We use the number test and Continuous Ranked Probability Score (CRPS) to measure the accuracy and precision of the predictions. We found that we need at least one mainshock and corresponding offspring to make reliable forecasting. And when we have more mainshocks in our data, our forecasting will be better. Besides, we also figure out what is needed to obtain a good posterior distribution for each parameter.
\end{abstract}

\section{Introduction}
\label{sec:intro}
Earthquake is one of the natural disasters that threaten human life and property safety. Compared with other natural disasters, the earthquake occurred suddenly, and the people in the disaster area could not escape in time. Therefore, research on earthquakes has never stopped. Although we still haven't realised the prediction of a future mainshock, which is the highest ideal of the research in earthquakes, good progress has been made in aftershock prediction. Reliable forecasting of aftershocks can help governments and rescuers better plan post-earthquake relief efforts. It can not only win more survival opportunities for the affected people but also better protect the lives of rescuers.\\

\noindent
Epidemic type aftershock sequence (ETAS) \citep{ogata1988statistical} is currently the most popular for modelling earthquake catalogues. It is a self-excited point process. New events can be generated by background rate or triggering functions of historical events. Besides, the new event increases the intensity in a short time period and spatial extent.\\

\noindent
Here are some methods for estimating the model parameters. Maximum likelihood estimation (MLE) is the most simple and direct method, but it cannot obtain the uncertainty of parameter estimation. Markov chain Monte Carlo (MCMC), as a Bayesian method, can get the posterior distribution of model parameters \citep{ross2021bayesian}. But the process usually takes a long time, and the high correlation between model parameters cannot guarantee that a good parameter posterior distribution is obtained. `inlabru' can significantly reduce the model fitting time by approximating the posterior distribution and using Integrated Nested Laplace Approximation (INLA \citep{rue2017bayesian}) for Bayesian inference \citep{serafini2023approximation} \citep{naylor2023bayesian}. Besides,  INLA can do an excellent job if the dependence between covariance parameters and between latent variables is approximately linear, as it estimates a Gaussian approximation internally. \\

\noindent
In this article, we use the R package `inlabru' to implement some experiments on the Aquila earthquake, Amatrice earthquake and synthetic catalogues based on them. We aim to figure out three things. The first one is how the posterior distribution changes when fixing a model parameter to the true value or a wrong value. During this, we also find the model fitting time reduces significantly when we fix some parameters. The second one compares the inter-event-time distribution of the real data and synthetic catalogues. Through this, we can know whether the synthetic catalogues generated by the model are reasonable. And we also try to figure out how each parameter impacts the inter-event-time distribution. The last one is the goodness of forecasting of the model. We calculate the number test (N-test) score \citep{schorlemmer2007earthquake} and Continuous Ranked Probability Score (CRPS) \citep{hersbach2000decomposition} for multiple periods. At the same time, we explore what is needed to make a good forecast.\\

\noindent
The article is structured as follows: Section \ref{sec:methods} describes the ETAS model, the rough theory behind Bayesian inference of the model parameters using `inlabru', and introduces the definition of number test and CRPS. Section \ref{sec:results} shows the results of our three experiments, including the posterior distribution when fixing some parameters to the true value or wrong value, the inter-event-time distribution of real data and synthetic catalogues, and the accuracy and precision of the forecasting in multiple periods. Section \ref{sec:discussion} discusses the limitation of our experiments, and section \ref{sec:conclusion} concludes our work and introduces our future work.

\newpage

\section{Theory and Methods}
\label{sec:methods}

\subsection{The temporal ETAS model}
The epidemic type aftershock sequence (ETAS) model is based on the Hawkes process model \citep{hawkes1971point} \citep{hawkes1971spectra}, a counting process. The general form of the intensity of the Hawkes process model is defined by
\[
\lambda_{Hawkes} (\mathbf{x}|\mathcal{H}_t) = \mu(\mathbf{x}) + \sum_{\mathbf{x}_h\in \mathbb{H}_t} g(\mathbf{x}, \mathbf{x}_h).
\]
The former term is the background rate, and the latter is the sum of the triggering function. $\mathbf{x}$ is an event, and $\mathcal{H}_t$ is the set of the events happened before $\mathbf{x}$. An event can be either generated from the background rate, having nothing to do with any history event, or generated from the triggering function of a history event. Thus, the intensity of any event is affected by all historical events, and any event influences the intensity of future events. Such a setting is very suitable for describing earthquake events.\\

\noindent
In the context of the earthquake, $\mathbf{x} = (t, \mathbf{s}, m)\in T\times W\times M$, where $t$ is the occurring time, $\mathbf{s} = (x, y)$ stands for the occurring position, and $m$ represents the magnitude. Assume $\mu$ is stable over time at a fixed location, and $g(\mathbf{x}, \mathbf{x}_h)$ can be factorised as $g_t(t-t_h)g_\mathbf{s}(\mathbf{s}-\mathbf{s}_h)g_m(m)$. We can obtain the temporal Hawkes process,
\[
\begin{aligned}
	\lambda_{Hawkes}(t|\mathcal{H}_t) &= \int_W [\mu(\textbf{s}) +\sum_{\mathbf{x}_h\in \mathcal{H}_t}g_t(t- t_h)g_\mathbf{s}(\mathbf{s}-\mathbf{s}_h)g_m(m_h)]d\textbf{s}\\
	&= \int_W \mu(\textbf{s})d\textbf{s} + \sum_{\mathbf{x}_h\in \mathcal{H}_t}\int_W g_\mathbf{s}(\mathbf{s}-\mathbf{s}_h)d\textbf{s}g_t(t- t_h)g_m(m_h).
\end{aligned}
\]
As $\int_W \mu(\textbf{s})d\textbf{s}$ and $\int_W g_\mathbf{s}(\mathbf{s}-\mathbf{s}_h)d\textbf{s}$ are constant, the intensity of temporal Hawkes process can be simplified to 
\[
	\lambda_{Hawkes}(t,m|\mathcal{H}_t) = \mu + \sum_{(t_h,m_h)\in \mathcal{H}_t}g_t(t- t_h)g_m(m_h),
\]
where $g_t()$ is the scaled version of the old one, we use the same notation.\\

\noindent
The temporal ETAS model is an extended Hawkes process, marked by the magnitude distribution.
\[
\begin{aligned}
	\lambda_{ETAS}(t,m|\mathcal{H}_t) &= (\mu + \sum_{(t_h,m_h)\in \mathcal{H}_t}g_t(t- t_h)g_m(m_h))\pi(m)\\
	&=(\mu + \sum_{(t_h,m_h)\in \mathcal{H}_t}Ke^{\alpha(m_h-M_0)}(\frac{t-t_h}{c}+1)^{-p})\pi(m).
\end{aligned}
\]
K is productivity, scale the intensity of the triggering function. $\alpha$ is magnitude scaling, controlling the influence of the magnitude of the historical event. When $\alpha$ is big, The difference between the impact of a large-magnitude event and a small-magnitude event will be greater. c and p are time offset and aftershocks decay, determining how fast the influence slows down over time. Small c and large p will lead to a faster decrease in the intensity of the triggering function. \\

\noindent
$\pi(m)$ follows the Gutenberg-Richter Law \citep{wesnousky1994gutenberg}. The Law describes the relationship between the amount and magnitude of events, which is
\[
\log_{10}N(m) = a-bm.
\]
Equivalently,
\[
m\sim \text{Exp}(\beta), \quad\text{where } \beta = \frac{2}{3}b
\]

\subsection{Hawkes process log-likelihood approximation for inlabru}
We first approximate the log-likelihood of the Hawkes process model and then use the Integrated Nested Laplace Approximation (INLA) to implement Bayesian inference of the parameters. These procedures are integrated into the `inlabru' package. \\

\noindent
The log-likelihood of the Hawkes process model is
\[
\mathcal{L}(\theta|\mathcal{H}) = -\Lambda(T_1, T_2) + \sum_{(t_i, m_i) \in \mathcal{H}} \log\lambda (t_i|\mathcal{H}_{t_i})
\]
The first term is the integral of the intensity function, which is the expectation of the total number of events.
\[
\begin{aligned}
	\Lambda(T_1, T_2) &= \int_{T_1}^{T_2}\mu dt + \sum_{(t_i,m_i)\in \mathcal{H}} \int_{T_1}^{T_2} Ke^{\alpha(m_i-M_0)}(\frac{t-t_i}{c}+1)^{-p}\mathbb{I}(t>t_i)dt \\
	&= \Lambda_0(T_1, T_2) + \sum_{(t_i,m_i)\in \mathcal{H}}\Lambda_i(T_1, T_2).
\end{aligned}
\]
To improve the accuracy of the second term in the linear approximation later, we divide the interval of the integral into some small bins. Since the value of the triggering function decreases more and more slowly with the increase of t and converges to 0, we set narrower bins on the left side of the integral to improve accuracy and set wider bins on the right side of the integral to reduce unnecessary calculations. The bins developing strategy is as below,
\[
t_i, t_i+\Delta, t_i+\Delta(1+\delta), t_i+\Delta(1+\delta)^2,\cdots,t_i+\Delta(1+\delta)^{n_i}, T_2,
\]
where $\Delta$ control the length of the first bin and $\delta$ adjust the speed of bins becoming longer. For event $(t_i, m_i)$, denoting the number of bins is $B_i$, and the endpoints of bins are $t_0^{(b_i)}, t_1^{(b_i)}, \cdots, t_{B_i-1}^{(b_i)}$. Then the log-likelihood of the Hawkes process is 
\[
	\mathcal{L}(\theta|\mathcal{H}) = -\Lambda_0(T_1, T_2) - \sum_{(t_i,m_i)\in \mathcal{H}}\sum_{j=0}^{B_i-1}\Lambda_i(t_j^{(b_i)}, t_{j+1}^{(b_i)}) + \sum_{(t_i, m_i) \in \mathcal{H}} \log\lambda (t_i|\mathcal{H}_{t_i}).
\]
The linear approximate log-likelihood is
\[
	\overline{\mathcal{L}}(\theta|\mathcal{H}) = -\exp\{\overline{\log \Lambda_0}(T_1, T_2)\} - \sum_{(t_i,m_i)\in \mathcal{H}}\sum_{j=0}^{B_i-1} \exp\{\overline{\log \Lambda_i}(t_j^{(b_i)}, t_{j+1}^{(b_i)}) \} + \sum_{(t_i, m_i) \in \mathcal{H}} \overline{\log\lambda} (t_i|\mathcal{H}_{t_i}).
\]
$\overline{\log f}(\theta)$ is linearly approximated by the truncated Taylor series expanded at $\theta^*$.
\[
\overline{\log f}(\boldsymbol{\theta}) \approx \log f(\boldsymbol{\theta}^*) + \frac{1}{f(\boldsymbol{\theta}^*)}\sum_{j=1}^m (\theta_j - \theta_j^*)\frac{\partial }{\partial\theta_j}f(\boldsymbol{\theta})|_{\boldsymbol{\theta}=\boldsymbol{\theta}^*}.
\]

\subsection{The prior for the model}
From previous work. We know that `inlabru' is robust with the choice of the prior \citep{serafini2023approximation} and initial value \citep{naylor2023bayesian}. This report uses the same priors and initial values for different experiments. The priors are
\[
\mu \sim \text{Gamma}(0.3,0.6),\quad K, \alpha, c\sim U(0,10), \quad p\sim U(1,10).
\]
The initial values are
\[
\mu_0 = 0.5, K_0 = 0.1, \alpha _0 = 1, c_0 = 0.1, p_0 = 1.1.
\]

\noindent
Since INLA only support Gaussian prior for the parameters, we use the Inverse Probability Integral Transform (Inverse PIT) method to convert the sample from the Gaussian distribution to the sample from the target distribution we want. Denote $\theta$ follows $N(0,1)$ and $F_{Y}$ is the (Cumulative Distribution Function) CDF of the target distribution, then
\[
\eta(\theta) = F_{Y}^{-1}(\Phi(\theta)).
\]
$\eta(\theta)$ follows the target distribution, and here the target distribution is the prior we set.

\subsection{Fitting the model}
In the process of fitting the model, we start with the initial parameter $\boldsymbol{\theta}_0$, and use it as the point to obtain the Taylor series as the approximated log-likelihood function. Then we sum it with the log-prior and calculate the mode of the posterior distribution, denoting it $\boldsymbol{\theta}_1^*$. We update the parameter using line search by $\boldsymbol{\theta}^* = \alpha  \boldsymbol{\theta}_0 + (1-\alpha) \boldsymbol{\theta}_1^*$. We use $\boldsymbol{\theta}^*$ to replace $\boldsymbol{\theta}_0$ and repeat the process above until the difference between $\boldsymbol{\theta}_0$ and $\boldsymbol{\theta}^*$ is less than 1\%.

\subsection{The inter-event-time distribution}
\label{sec:interevent}
Interval time is the time difference between adjacent events. The author in \citep{molchan2005interevent} said that if the earthquake event follows a homogeneous Poisson point process, then the inter-event-time follows An Exponential distribution $\text{Exp}(\alpha)$. We know that 
\[
x \sim \text{Exp}(\alpha) \Rightarrow \frac{x}{\overline{x}} \sim \text{Exp}(1).
\]
Thus we normalise the inter-event-time with its mean and compare its CDF with that of Exp(1) to check the degree of clustering.

\subsection{Number test}
The number test is used to test the accuracy of the number of predicted events. We simulate the catalogues for $m$ times and count the events in a specific period, denoting $N_1, N_2, \cdots, N_m$. Then we have a distribution of the number of events in that period. As there are usually some outliers in the predictions, we typically use the median number as the prediction. And we hope the forecast can be as close to the true value as possible. The number test score can help us know the quantile position of the true value in the distribution of predictions, which is defined by
\[
\delta_1 = \frac{|\{N_j|N_j\geq N_{obs}, j=1,\cdots,m\}|}{m},\quad \delta_2 = \frac{|\{N_j|N_j\leq N_{obs}, j=1,\cdots,m\}|}{m}.
\]

\noindent
We use $\delta_2$ in this report. $\delta_2=0.5$ indicates that our prediction distribution is centred at the true value, while $\delta_2>0.5$ indicates underprediction, and  $\delta_2<0.5$ indicates overprediction.

\subsection{Continuous Ranked Probability Score}
Continuous Ranked Probability Score (CRPS) measure the accuracy and precision of the number of predicted events. While we hope that the median of the distribution is close to the true value, we also hope that the variance of the distribution is slight (the distribution is very narrow) so that we have great confidence in the prediction and CRPS can do it. The CRPS is given by
\[
S(F, N_{true}) = \sum_{k=0}^{\infty}(F(k)-\mathbb{I}(N_{true}\leq k))^2.
\]
 We use the empirical Cumulative Distribution Function (eCDF) $\hat F(x)$ to approximate the CDF $F(x)$, which can be calculated by
\[
\hat F(x) = \frac{|\{N_j|N_j \leq x\}|}{|\{N_j\}|},\quad x=0,1,2\cdots.
\]
As we know that $\hat F(x) = 0$ when $x<\min(N_j)$ and $\hat F(x) = 1$ when $x\geq\max(N_j)$, we can only sum up from $\min(N_j)$ to $\max(N_j)$ when calculating the CRPS instead of from 0 to $\infty$.\\

\noindent
When the CRPS is small, it indicates that our prediction distribution is centred around the true value with a very small variance.

\newpage

\section{Results}
\label{sec:results}
\subsection{Influence of fixing one parameter when fitting the model}
\label{sec:fix}

To figure out how the five parameters in the model influence the posterior distribution of each other, we fix one parameter at a time to the true or wrong value, fit the model and compare the posterior distribution of other parameters. We fix a parameter by using a same kind of prior with the mean equal to the specific value and an extremely small variance. For Gamma distribution, we use $\text{Gamma}(value\times \epsilon, \epsilon)$, and for Uniform distribution, we use $\text{U}(value-\epsilon, value+\epsilon)$.\\

\noindent
We implement the experiments on synthetic catalogues so that we know the true value of the parameters. The parameters to generate the synthetic catalogues come from the posterior mode of the model fitted with the Aquila earthquake catalogue. And the mainshock of Aquila is used as a historical event. In this case, the synthetic catalogues would behave like the Aquila earthquake. We implement the experiments on three different synthetic catalogues to ensure that the results do not happen by chance.\\

\begin{figure}[H]
	\includegraphics[width=\textwidth]{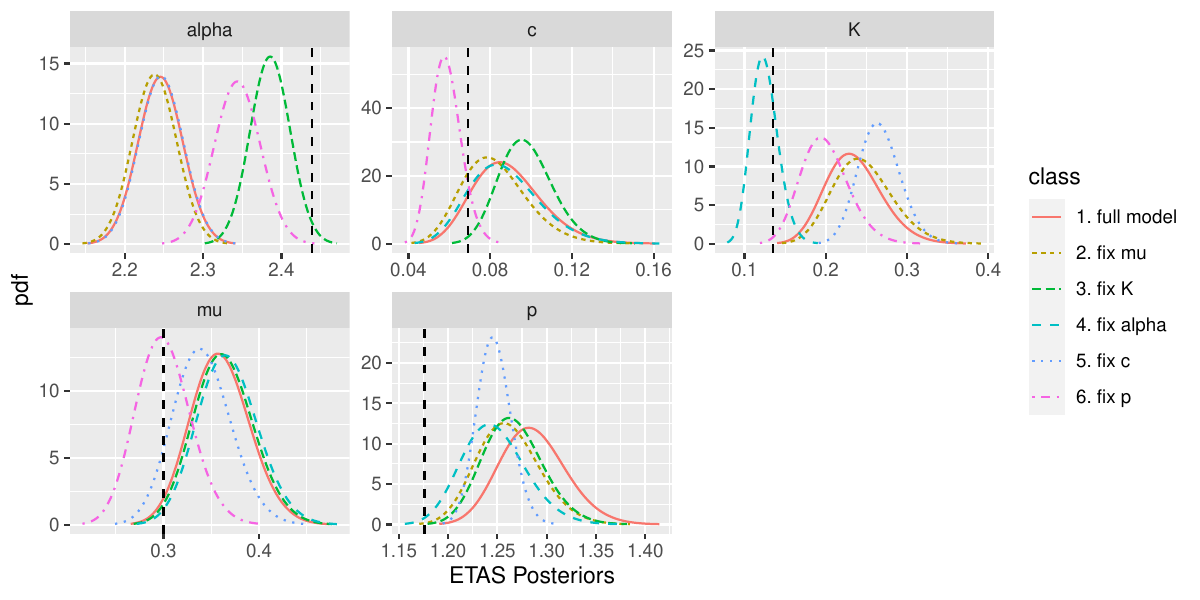}
	\caption{ETAS posteriors distribution on synthetic catalogue 1 when fixing one parameter to the true value}
	\label{fig:fix_para_cat1}
\end{figure}

\begin{figure}[H]
	\includegraphics[width=\textwidth]{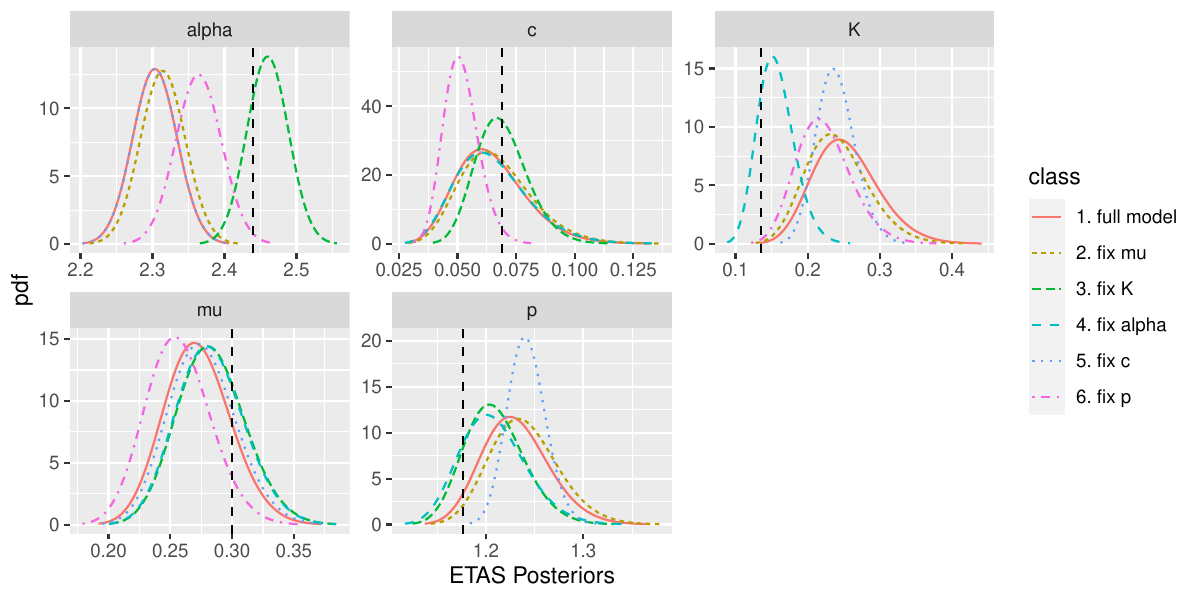}
	\caption{ETAS posteriors distribution on synthetic catalogue 2 when fixing one parameter to the true value}
	\label{fig:fix_para_cat2}
\end{figure}

\begin{figure}[H]
	\includegraphics[width=\textwidth]{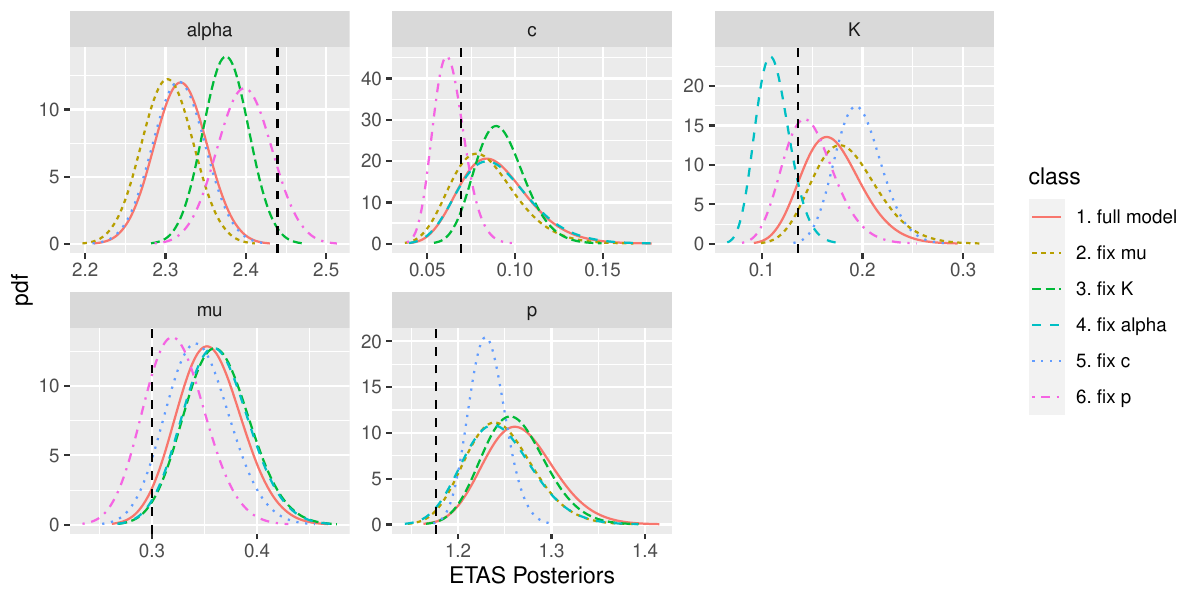}
	\caption{ETAS posteriors distribution on synthetic catalogue 3 when fixing one parameter to the true value}
	\label{fig:fix_para_cat3}
\end{figure}

\noindent
 \textbf{Figure} \ref{fig:fix_para_cat1}, \textbf{Figure} \ref{fig:fix_para_cat2}, \textbf{Figure} \ref{fig:fix_para_cat3} show the posterior distribution when we fix one parameter to its true value. From these figures, we can observe that:
\begin{itemize}
	\item $\alpha$ and $K$ have a clear mutual influence relationship. When we fix $K$ to the true value, the mean of the posterior distribution of $\alpha$ is closer to the true value, and the variance is smaller. Samely, when we fix $\alpha$ to the true value, the posterior mean of $K$ is closer to the true value, and the posterior variance is smaller, indicating we have more confidence in the posterior mean.
	\item There is also an apparent mutual influence relationship between $c$ and $p$. The posterior variance of $p$ is smaller when we fix $c$, and the posterior variance of $c$ is smaller when we fix $p$. Still, their posterior mean is not necessarily closer to the true value when we fix the other one.
	\item When we fix $p$, the posterior mean of $\alpha$ is closer to the true value.
	\item When we fix $\mu$, the posterior distributions of other parameters do not change. At the same time, fixing other parameters will not affect the posterior distribution of $\mu$.
	
\end{itemize}

\noindent During the experiment, we notice that the model fitting time decreases significantly when fixing some parameters. From the \textbf{table} \ref{table:1}, we can observe that when fixing $K$ or $\alpha$, the model fitting time decrease significantly, especially for $\alpha$. Therefore, if we know the value of $\alpha$ or $K$ in advance or have some information about them, we can save much time fitting the model.

\begin{table}[H]
\centering
\begin{tabular}{c c c c c c c} 
 \hline
  & full model & fix mu & \textbf{fix K} & \textbf{fix alpha} & fix c & fix p\\
 \hline
 catalogue 1 & 41.73 & 41.56 & \textbf{26.00} & \textbf{18.25} & 42.76 & 36.85 \\ 
 catalogue 2 & 46.51 & 45.84 & \textbf{20.95} & \textbf{7.75} &  45.97 & 43.27\\
 catalogue 3 & 44.92 & 43.98 & \textbf{21.47} & \textbf{15.15} & 45.12 & 42.77 \\
 \hline
\end{tabular}
\caption{Model fitting time when fixing one parameter to the true value}
\label{table:1}
\end{table}

\begin{figure}[H]
	\centering
	\includegraphics[width=0.7\textwidth]{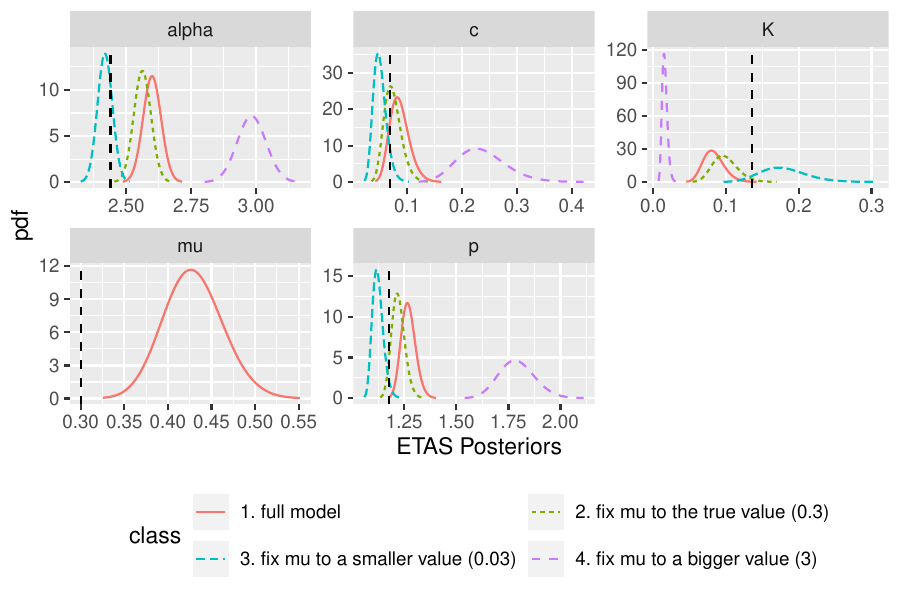}
	\caption{ETAS posteriors distribution when fixing $\mu$ to a wrong value}
	\label{fig:fix_para_cat4}
\end{figure}

\begin{figure}[H]
	\centering
	\includegraphics[width=0.7\textwidth]{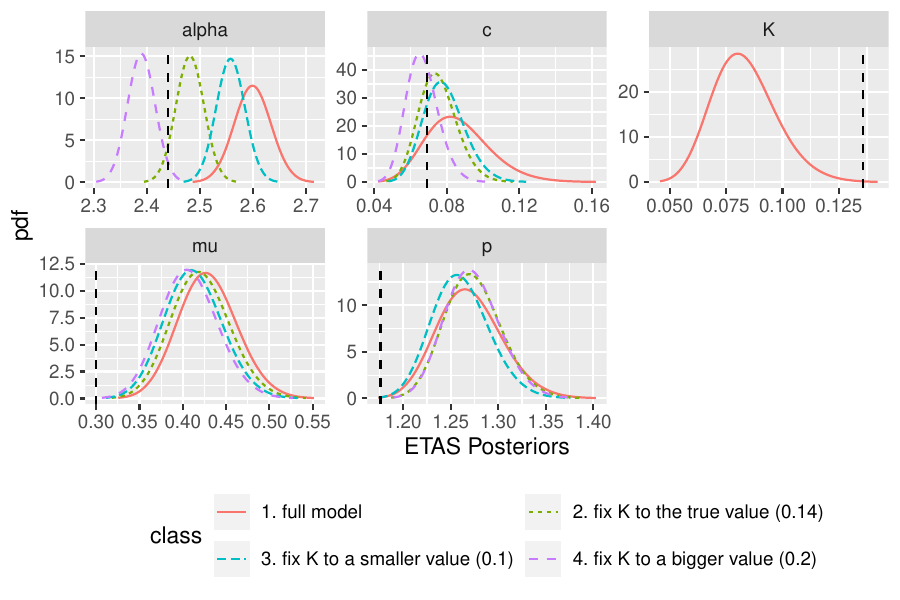}
	\caption{ETAS posteriors distribution when fixing $K$ to a wrong value}
	\label{fig:fix_para_cat5}
\end{figure}

\begin{figure}[H]
	\centering
	\includegraphics[width=0.7\textwidth]{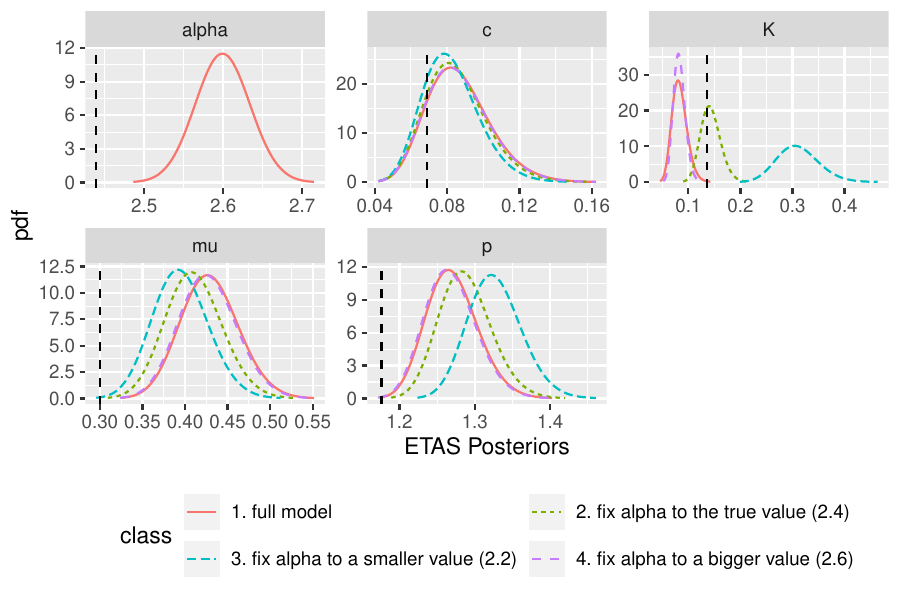}
	\caption{ETAS posteriors distribution when fixing $\alpha$ to a wrong value}
	\label{fig:fix_para_cat6}
\end{figure}

\begin{figure}[H]
	\centering
	\includegraphics[width=0.7\textwidth]{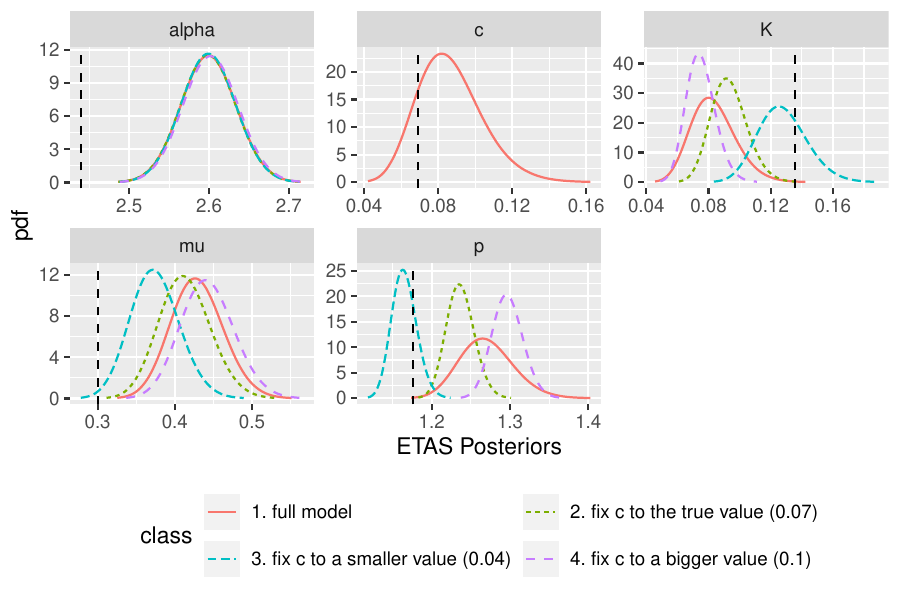}
	\caption{ETAS posteriors distribution when fixing $c$ to a wrong value}
	\label{fig:fix_para_cat7}
\end{figure}

\begin{figure}[H]
	\centering
	\includegraphics[width=0.7\textwidth]{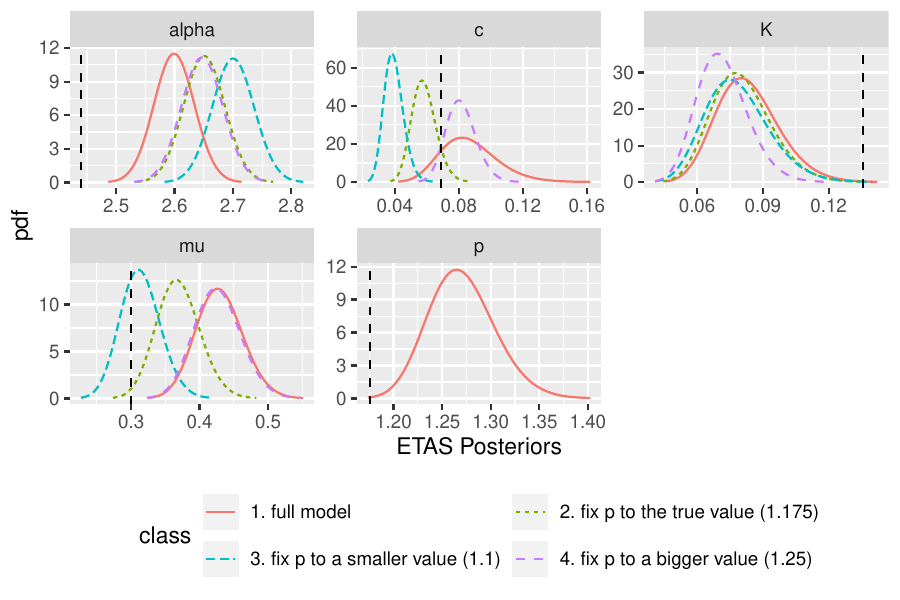}
	\caption{ETAS posteriors distribution when fixing $p$ to a wrong value}
	\label{fig:fix_para_cat8}
\end{figure}

\noindent
The above experiment reflects the influence of knowing the correct information of the parameters on other parameters' posterior distributions. Since most of the time, fixing one parameter to the true value does not affect the posterior distribution of other parameters, especially $\mu$. We fix one parameter to a value smaller or larger than the true value to further explore the interaction between parameters. \textbf{Figure} \ref{fig:fix_para_cat4}, \textbf{Figure} \ref{fig:fix_para_cat5}, \textbf{Figure} \ref{fig:fix_para_cat6}, \textbf{Figure} \ref{fig:fix_para_cat7}, \textbf{Figure} \ref{fig:fix_para_cat8} show the posterior distribution of them. \\

\noindent
In previous experiments, we concluded that fixing $\mu$ to the true value does not affect the posterior distributions of the other parameters. But from \textbf{Figure} \ref{fig:fix_para_cat4}, it can be seen that fixing $\mu$ to the wrong value affects the posterior distribution of the other four parameters. 
Fixing $\mu$ to larger values causes the posterior distributions of $\alpha$, $c$ and $p$ to move to the right, and that of K to move to the left. Conversely, fixing $\mu$ to a smaller value will result in a leftward moving of $\alpha$, $c$, and $p$; and a rightward moving of $K$. The value of $\mu$ will determine the dominance of the triggering function in the intensity function of the Hawkes process. And larger $\mu$ has more influence than smaller $\mu$.

\noindent
Previous experiments have shown that $\alpha$ and $K$ interact, and the results in \textbf{Figures} \ref{fig:fix_para_cat5} and \ref{fig:fix_para_cat6} also support this phenomenon. When we fix one of the variables to a value larger than the true value, the posterior distribution of the other variable will be smaller. When we fix the value of one of the variables to be smaller than the true value, the posterior distribution of the other variable will be bigger. Furthermore, fixing $\alpha$ and $K$ to wrong values does not significantly affect the posterior distributions of the other parameters.

\noindent
From \textbf{Figure} \ref{fig:fix_para_cat7} and \textbf{Figure} \ref{fig:fix_para_cat8}, we can observe that fixing $c$ to a wrong value does not affect the posterior distribution of $\alpha$, and fixing $p$ to a wrong value has no effect on the posterior distribution of $K$.

\subsection{The inter-event-time distribution of real data and synthetic catalogues}
In \textbf{Section} \ref{sec:interevent}, we know that if the occurring time of the earthquake event follows homogeneous Poisson distribution, the normalised inter-event-time follows Exp(1). In this section, we compare the inter-event-time distribution of Aquila and synthetic catalogues. The synthetic catalogues are generated using the same parameters and history event as in \textbf{Section} \ref{sec:fix}, the posterior mode of the model fitted with the Aquila catalogue, and the mainshock of Aquila. Then we calculate their inter-event-time and normalised them with their mean separately.\\

\begin{figure}[H]
	\centering
	\includegraphics[width=0.6\textwidth]{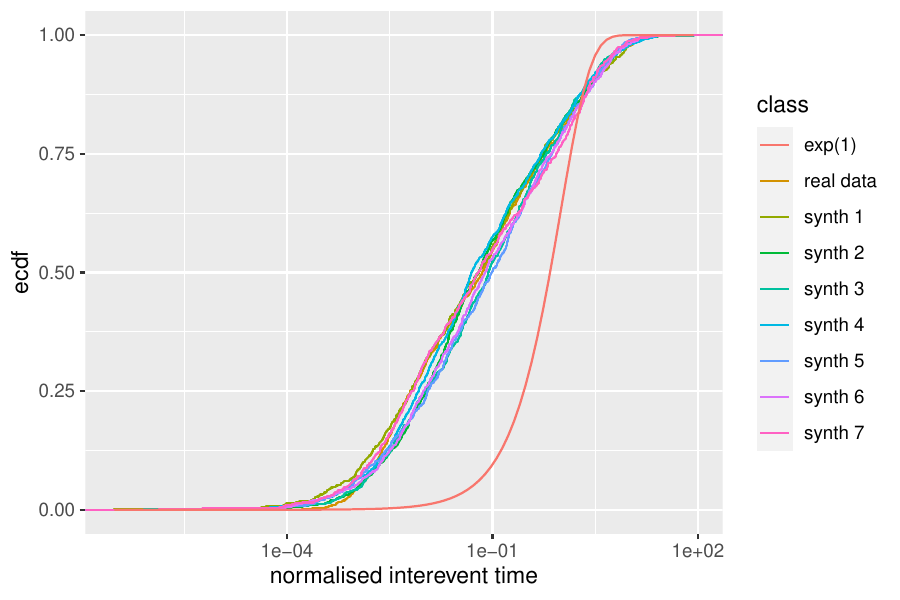}
	\caption{Empirical cumulative distribution function of real data and synthetic catalogues}
	\label{fig:inte_time1}
\end{figure}

\noindent
\textbf{Figure} \ref{fig:inte_time1} shows the empirical cumulative distribution function (eCDF) of the normalised inter-event-time of Aquila and synthetic catalogues. We can observe that the eCDFs of real data and synthetic catalogues are basically consistent. Compared with the eCDF of Exp(1), here are more short and long inter-event-times, indicating clustering in the real data and synthetic catalogues.

\begin{figure}[H]
	\centering
	\includegraphics[width=\textwidth]{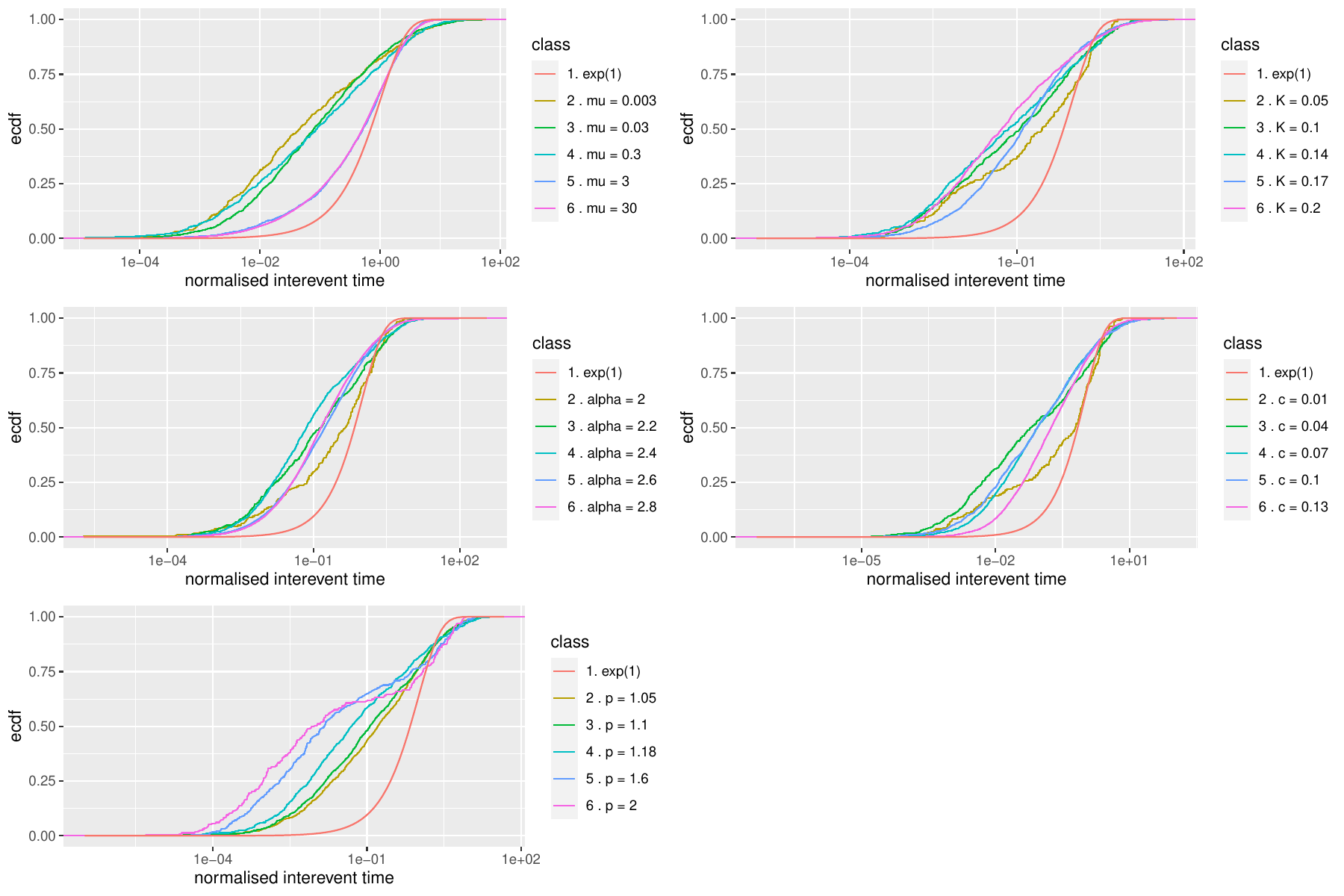}
	\caption{Empirical cumulative distribution function of synthetic catalogues when changing the value of one parameter}
	\label{fig:inte_time2}
\end{figure}

\noindent
We then changed one parameter at a time to different values to see how each parameter affects the normalised inter-event-time distribution. From \textbf{Figure} \ref{fig:inte_time2}, we can observe that $\mu$ and $p$ significantly affect the normalised inter-event-time distribution. As $\mu$ increases, the eCDF of the normalised inter-event-time gets closer and closer to Exp(1). The background rate $\mu$ dominates the ETAS functions, and the events approximately follow a homogeneous Poisson distribution. There are still many short intervals here, which the aftershocks brought by the mainshock may cause. As $p$ increases, the short inter-event-time grows in number significantly. The increase of $p$ makes the triggering function decrease faster over time, so that the aftershocks are closer to the mainshock.

\subsection{Impact of the number of historical mainshocks on forecasting}
We use the mainshock as a separation point, and we use the events preceding the mainshock to train the model and as historical events to predict the events after the mainshock. We include the mainshock (separation point) in historical events but not when training the model. Because when we include the mainshock for training the model, the mainshock at the end will behave like an outlier, thus affecting the training of the model.\\

\begin{figure}[H]
	\centering
	\includegraphics[width=0.5\textwidth]{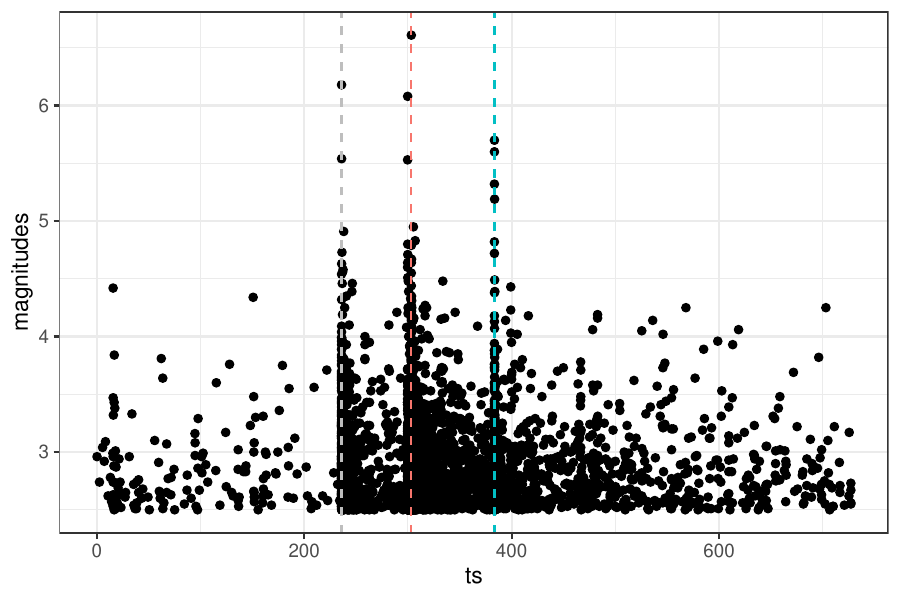}
	\caption{Amatrice earthquake catalogue}
	\label{fig:amatrice}
\end{figure}

\noindent
The catalogue we use is the Amatrice earthquake from 2016 to 2017 in Italy, shown in \textbf{Figure} \ref{fig:amatrice}. The Amatrice earthquake has three mainshocks. We made predictions for all three, but the first one failed. The reason will be explained later.

\begin{figure}[H]
	\centering
	\includegraphics[width=0.7\textwidth]{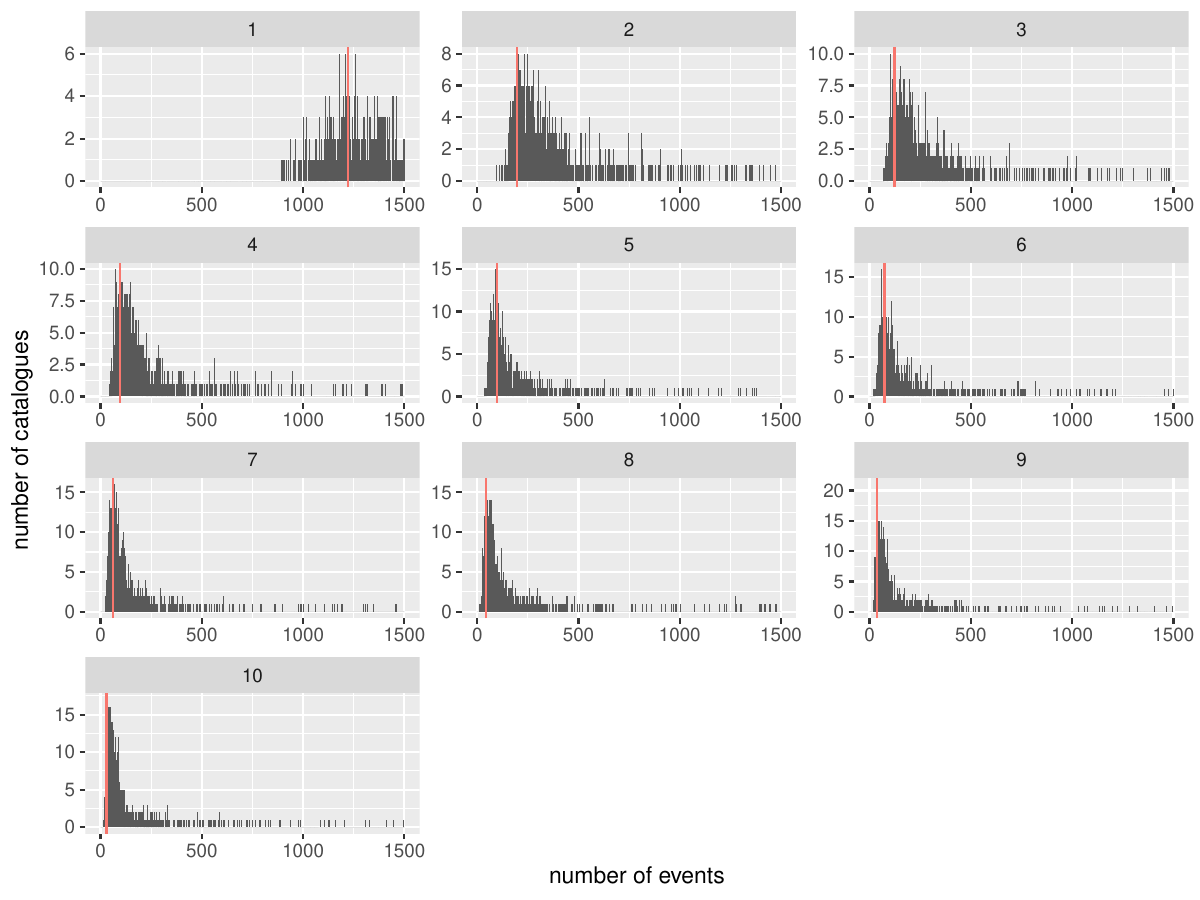}
	\caption{Forecasting distribution for the events after the second mainshock}
	\label{fig:ntest1}
\end{figure}

\begin{figure}[H]
	\centering
	\includegraphics[width=0.7\textwidth]{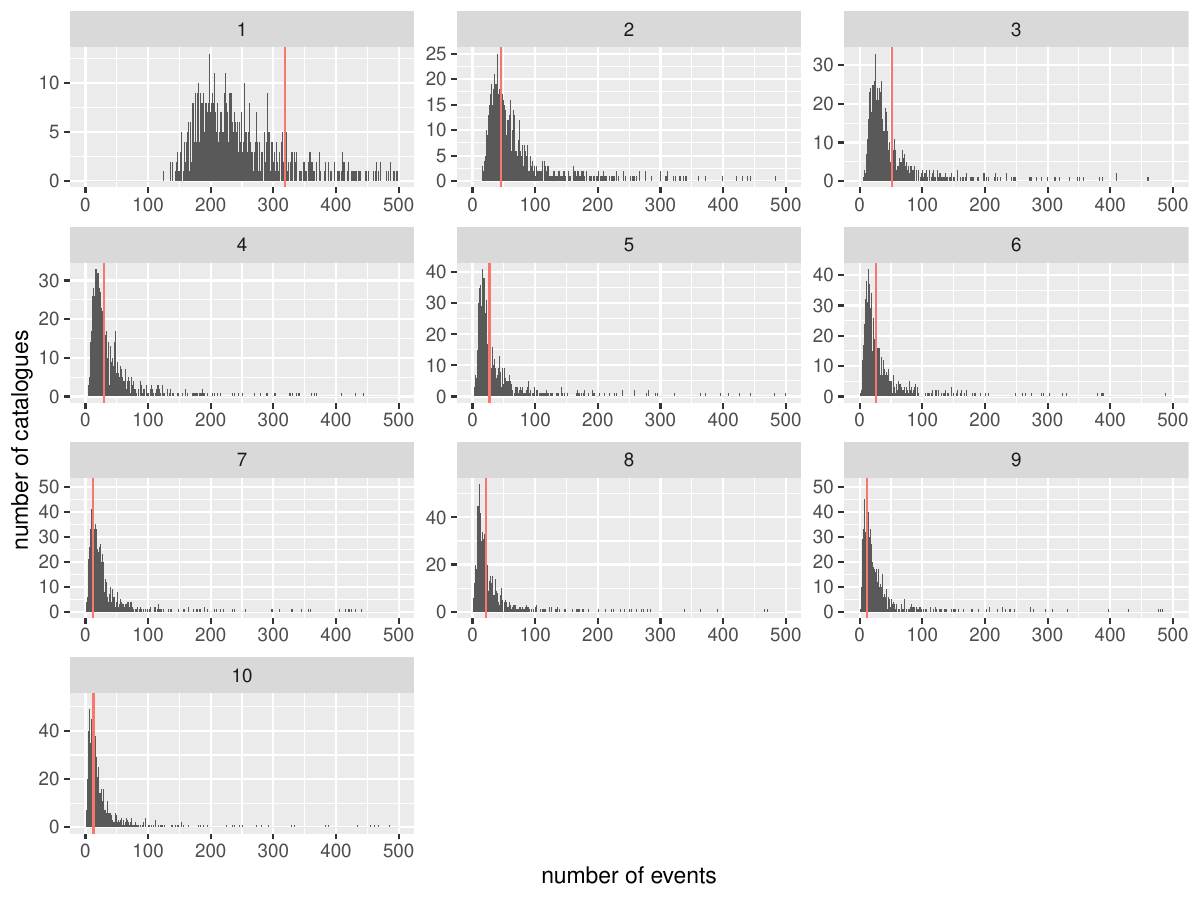}
	\caption{Forecasting distribution for the events after the third mainshock}
	\label{fig:ntest2}
\end{figure}

\noindent
\textbf{Figure} \ref{fig:ntest1} shows the results using the second mainshock as the separation point. We forecast ten weeks of earthquake events after the mainshock. The position of the black line indicates the number of events for each week, and the height indicates the number of catalogues. The red line represents the actual number of events. We can observe that during these weeks, the black line clusters around the red line, which shows that our forecast is good. \textbf{Figure} \ref{fig:ntest2} shows the results using the third mainshock as the separation point. The result is not bad either because the red line is within the black line.

\begin{figure}[H]
	\centering
	\includegraphics[width=0.6\textwidth]{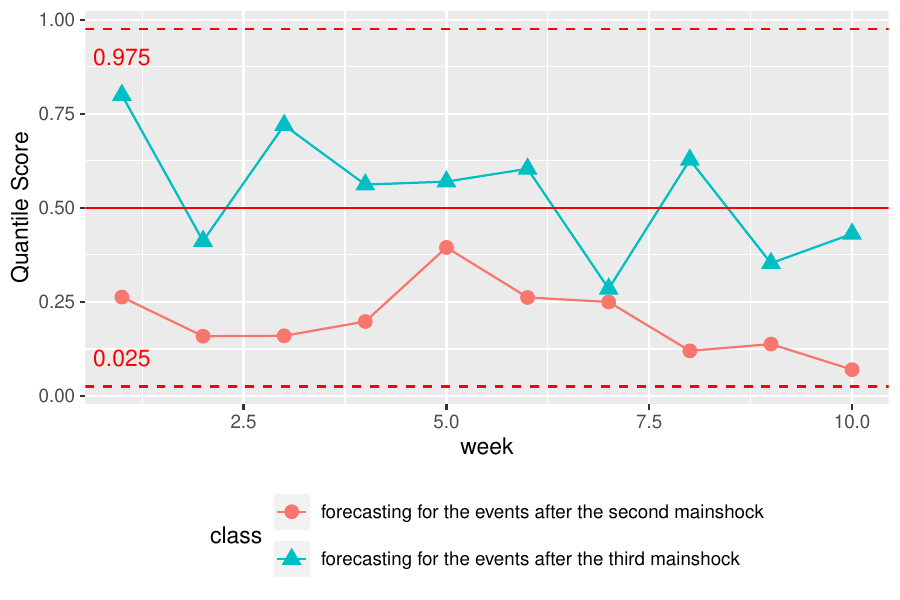}
	\caption{The result of number test for including one or two mainshocks and corresponding aftershocks}
	\label{fig:ntest}
\end{figure}

\noindent
\textbf{Figure} \ref{fig:ntest} shows the number test results for the two forecasts. We can observe that for two forecastings, the true values are within the 95\% symmetric credible interval of the predicted values, indicating good prediction. The forecasting for the events after the third mainshock performs better, as $\delta_2$ values in each week are around 0.5, indicating that the true values are in the centre of the predictions. The forecasting for the events after the second mainshock is overpredicted in all ten weeks.

\begin{figure}[H]
	\centering
	\includegraphics[width=0.6\textwidth]{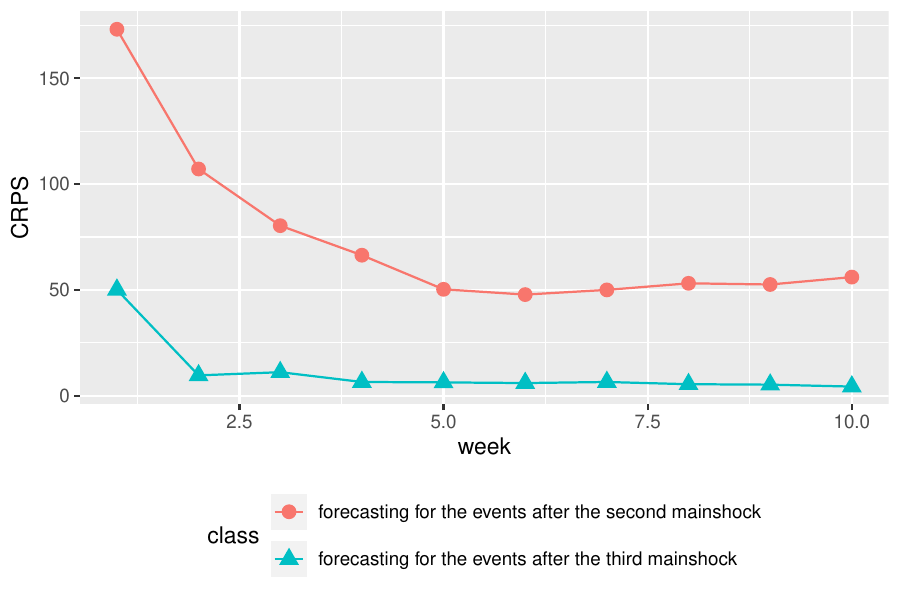}
	\caption{The result of Continuous Ranked Probability Score for including one or two mainshocks and corresponding aftershocks}
	\label{fig:crps}
\end{figure}

\noindent
From \textbf{Figure} \ref{fig:crps} shows the result of Continuous Ranked Probability Score (CRPS) for the two forecastings, we can observe that the forecasting for the events after the third mainshock is much better than the forecasting of events after the second mainshock.

\begin{figure}[H]
	\centering
	\includegraphics[width=0.7\textwidth]{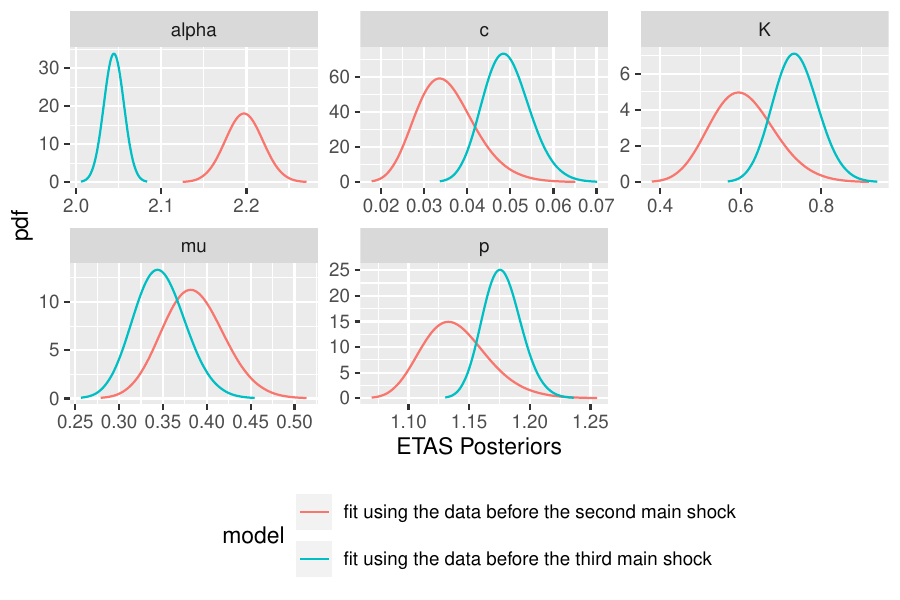}
	\caption{Posterior distributions when including more or less mainshocks and corresponding offspring}
	\label{fig:amatrice_post}
\end{figure}

\noindent
The magnitude distribution parameter's estimates for the two models are similar, 2.33 and 2.39. Thus the potential reason for the two forecastings' different performances should be the five parameters' posterior distributions. \textbf{Figure} \ref{fig:amatrice_post} shows the posterior distributions for both models. The red and blue lines represent the models fitted with the data before the second and third mainshocks, respectively. We can observe that there are overlaps for the posterior distribution of $c, K, \mu$, and $p$. But for $\alpha$, the posterior distributions of the two models are far apart, with no overlap. Therefore, when we fit the model by including more mainshocks and corresponding aftershocks, the posterior distribution for $\alpha$ gets better. Also, the posterior distribution of $\alpha$ is likely responsible for the predictive performance in this case. More experiments are needed to figure this out.

\begin{figure}[H]
	\centering
	\includegraphics[width=0.7\textwidth]{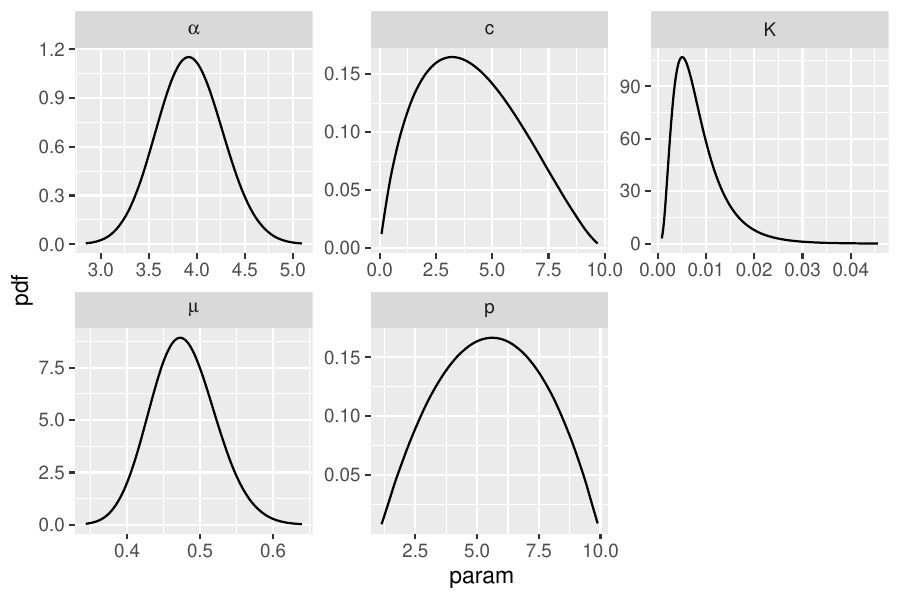}
	\caption{Posterior distribution when including no mainshock and corresponding offspring}
	\label{fig:amatrice_post3}
\end{figure}

\noindent
We also fit the model using the data before the first mainshock. In that case, we don't include any mainshock and offspring. The posterior distributions are shown in \textbf{Figure} \ref{fig:amatrice_post3}. We can observe that the posterior distributions of the parameters except for $\mu$ are terrible. 

\begin{figure}[H]
	\centering
	\includegraphics[width=0.7\textwidth]{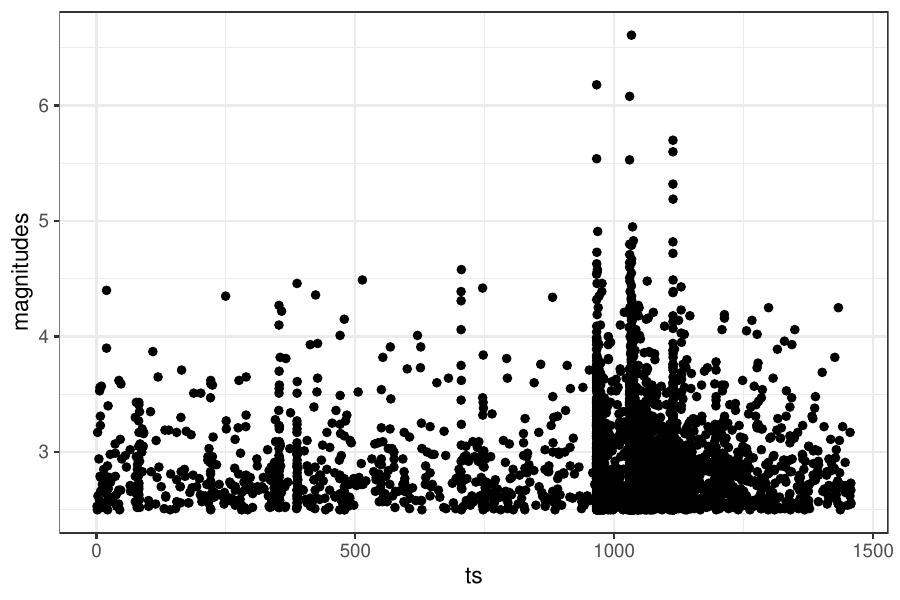}
	\caption{Amatrice earthquake catalogue (including data from two years ago)}
	\label{fig:amatrice2}
\end{figure}

\begin{figure}[H]
	\centering
	\includegraphics[width=0.7\textwidth]{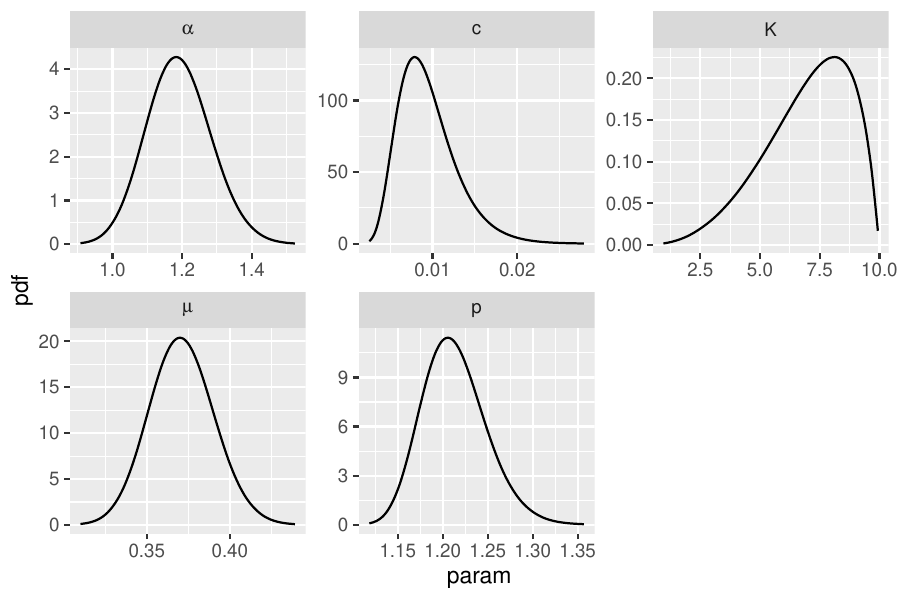}
	\caption{Posterior distribution when including no mainshock and corresponding offspring (including data from two years ago)}
	\label{fig:amatrice_post4}
\end{figure}

\noindent
Insufficient events may also cause a worse posterior distribution. To determine whether this causes it, we include the data from two years ago to fit the model. Similarly, we don't include any mainshock and corresponding offspring. The extra events were shown in \textbf{Figure} \ref{fig:amatrice2}, and the posterior distribution of the model was shown in \textbf{Figure} \ref{fig:amatrice_post4}. We can observe that the mode of $\mu$ becomes smaller, closer to those in \textbf{Figure} \ref{fig:amatrice_post}. The posterior distributions of $c$ and $p$ are much better than those in \textbf{Figure} \ref{fig:amatrice_post3}. The posterior for $\alpha$ and $K$ is still terrible. We need at least one mainshock and its offspring to estimate $\alpha$ and $K$. When more mainshocks and offspring are included, the estimation would be better.

\newpage

\section{Discussion}
\label{sec:discussion}
When studying the effect of fixing one parameter on the posterior distribution of other parameters, and the effect of a parameter on the inter-event-time distribution, the conclusions obtained are limited because the synthetic catalogue we use has only one mainshock. We can repeat the same experiment on synthetic catalogues with more mainshocks to see if the results are consistent or if some new insights emerge.\\

\noindent
The inter-event-time defined in this report is the time difference between adjacent events. We can further cluster events by mainshock and explore the distribution of inter-event-time within and between groups, as the authors in \citep{touati2009origin} did. \\

%\noindent
%
%
%\noindent
%When making predictions, we have found that including more mainshocks and corresponding aftershocks when training the model and in historical events improves the accuracy and precision of the predictions. When predicting aftershocks of a mainshock that occurred in a place, we need at least one past mainshock and the corresponding aftershock data in that place. Use them to train the model to obtain a more reasonable posterior distribution of $\alpha$ and $K$.\\

\noindent
As the number of main shocks increases, the fitting time of the model will be significantly longer, and it may even be difficult to converge. A parameter update strategy might solve this problem by first fitting the model on the previous mainshock and aftershock to obtain the posterior distribution, and then using the posterior distribution as the prior distribution to fit the next mainshock and aftershock.\\

\noindent
We can make predictions on the synthetic catalogues. This way, we know the true values of the parameters so that we can compare the quality of their posterior distributions. But besides the number of mainshocks, many other earthquake characteristics also affect the posterior distribution. Finding these characteristics and designing controlled experiments is crucial. 

\newpage

\section{Conclusion and future work}
\label{sec:conclusion}
In the first experiment, the prior information of $\alpha$ and $K$ is the most important among the five parameters. Knowing any of them can improve the accuracy and precision of the other's posterior distribution and significantly reduce the model's training time. When unsure whether the prior information of $\mu$ is accurate, the uninformative prior is more suitable for $\mu$. Because the correct prior information of $\mu$ will not make the posterior distribution of other parameters more accurate, but the wrong prior information of $\mu$ will affect the posterior distribution of other parameters.\\

\noindent
In the second experiment, the normalised inter-event-time distributions of the real data and the synthetic catalogues are basically consistent, indicating that the synthetic catalogues generated from the model are reasonable and our model can well describe the distribution of earthquake events. A larger $\mu$ will make the normalised inter-event-time distribution more similar to Exp(1), which means that the event is more like to follow a homogeneous Poisson distribution. A larger p increases the number of short inter-event-time because it makes aftershocks more concentrated.\\

\noindent
In the last experiment, the model has reliable predictive ability. We use the events before the mainshock to predict the events ten weeks after the mainshock, and the number of predicted events per week has good accuracy and precision. At least one historical mainshock and corresponding offspring are needed for reliable forecasting. The more historical mainshocks and corresponding aftershocks we have, the better our predictions of future aftershocks will be.\\

\noindent
During the last experiment, we also conclude what is needed to obtain a good posterior distribution. $\mu$ is the easiest to estimate; we can estimate $\mu$ with few events. $c$ and $p$ can be estimated in a quiet catalogue, but we need more earthquake events. $\alpha$ and $K$ are the hardest to estimate, we need at least one mainshock and corresponding aftershocks in the catalogue. When we have more mainshocks, the posterior of $\alpha$ gets significantly better. We believe that the posterior of $\alpha$ gets better as the number of mainshocks increases and even better when the mainshocks have different magnitudes.\\

\noindent
Our work can lead to a better understanding of the interactions between parameters and the impact of parameters on the inter-event-time distribution. More importantly, we summarise what is needed for making a reliable forecast or a better posterior distribution. In future work, we will conduct forecasting experiments on more synthetic catalogues. In this case, we know the real values of the parameters, and we can obtain a synthetic catalogue with more mainshocks. In addition, we will explore the impact of other earthquake characteristics on posterior distribution and forecasting. 

\newpage
\section{Acknowledgments}
Thanks to Prof. Finn Lindgren, Dr. Francesco Serani and Dr. Daniel Paulin for their guidance and
assistance with my thesis.

\newpage
{
\small
\bibliographystyle{plainnat}
\bibliography{neurips_2025.bib}

\begin{thebibliography}{12}
\providecommand{\natexlab}[1]{#1}
\providecommand{\url}[1]{\texttt{#1}}
\expandafter\ifx\csname urlstyle\endcsname\relax
  \providecommand{\doi}[1]{doi: #1}\else
  \providecommand{\doi}{doi: \begingroup \urlstyle{rm}\Url}\fi

\bibitem[Hawkes(1971{\natexlab{a}})]{hawkes1971point}
Alan~G Hawkes.
\newblock Point spectra of some mutually exciting point processes.
\newblock \emph{Journal of the Royal Statistical Society Series B: Statistical Methodology}, 33\penalty0 (3):\penalty0 438--443, 1971{\natexlab{a}}.

\bibitem[Hawkes(1971{\natexlab{b}})]{hawkes1971spectra}
Alan~G Hawkes.
\newblock Spectra of some self-exciting and mutually exciting point processes.
\newblock \emph{Biometrika}, 58\penalty0 (1):\penalty0 83--90, 1971{\natexlab{b}}.

\bibitem[Hersbach(2000)]{hersbach2000decomposition}
Hans Hersbach.
\newblock Decomposition of the continuous ranked probability score for ensemble prediction systems.
\newblock \emph{Weather and Forecasting}, 15\penalty0 (5):\penalty0 559--570, 2000.

\bibitem[Molchan(2005)]{molchan2005interevent}
G~Molchan.
\newblock Interevent time distribution in seismicity: a theoretical approach.
\newblock \emph{Pure and Applied Geophysics}, 162:\penalty0 1135--1150, 2005.

\bibitem[Naylor et~al.(2023)Naylor, Serafini, Lindgren, and Main]{naylor2023bayesian}
Mark Naylor, Francesco Serafini, Finn Lindgren, and Ian~G Main.
\newblock Bayesian modeling of the temporal evolution of seismicity using the etas. inlabru package.
\newblock \emph{Frontiers in Applied Mathematics and Statistics}, 9:\penalty0 1126759, 2023.

\bibitem[Ogata(1988)]{ogata1988statistical}
Yosihiko Ogata.
\newblock Statistical models for earthquake occurrences and residual analysis for point processes.
\newblock \emph{Journal of the American Statistical association}, 83\penalty0 (401):\penalty0 9--27, 1988.

\bibitem[Ross(2021)]{ross2021bayesian}
Gordon~J Ross.
\newblock Bayesian estimation of the etas model for earthquake occurrences.
\newblock \emph{Bulletin of the Seismological Society of America}, 111\penalty0 (3):\penalty0 1473--1480, 2021.

\bibitem[Rue et~al.(2017)Rue, Riebler, S{\o}rbye, Illian, Simpson, and Lindgren]{rue2017bayesian}
H{\aa}vard Rue, Andrea Riebler, Sigrunn~H S{\o}rbye, Janine~B Illian, Daniel~P Simpson, and Finn~K Lindgren.
\newblock Bayesian computing with inla: a review.
\newblock \emph{Annual Review of Statistics and Its Application}, 4:\penalty0 395--421, 2017.

\bibitem[Schorlemmer et~al.(2007)Schorlemmer, Gerstenberger, Wiemer, Jackson, and Rhoades]{schorlemmer2007earthquake}
Danijel Schorlemmer, MC~Gerstenberger, S~Wiemer, DD~Jackson, and DA~Rhoades.
\newblock Earthquake likelihood model testing.
\newblock \emph{Seismological Research Letters}, 78\penalty0 (1):\penalty0 17--29, 2007.

\bibitem[Serafini et~al.(2023)Serafini, Lindgren, and Naylor]{serafini2023approximation}
Francesco Serafini, Finn Lindgren, and Mark Naylor.
\newblock Approximation of bayesian hawkes process with inlabru.
\newblock \emph{Environmetrics}, page e2798, 2023.

\bibitem[Touati et~al.(2009)Touati, Naylor, and Main]{touati2009origin}
Sarah Touati, Mark Naylor, and Ian~G Main.
\newblock Origin and nonuniversality of the earthquake interevent time distribution.
\newblock \emph{Physical Review Letters}, 102\penalty0 (16):\penalty0 168501, 2009.

\bibitem[Wesnousky(1994)]{wesnousky1994gutenberg}
Steven~G Wesnousky.
\newblock The gutenberg-richter or characteristic earthquake distribution, which is it?
\newblock \emph{Bulletin of the Seismological Society of America}, 84\penalty0 (6):\penalty0 1940--1959, 1994.

\end{thebibliography}
}

\end{document}